\documentclass[twocolumn,narrowdisplay]{elsart3p}

\usepackage{graphicx,amssymb}

\journal{Synthetic Metals}

\begin{document}
\begin{frontmatter}

\title{From valence bond solid to unconventional superconductivity in the organic charge-transfer solids\thanksref{grants}}
\thanks[grants]{Supported by DOE Grant
No. DE-FG02-06ER46315 and NSF Grant No. DMR-0705163.}

\author[1]{S. Mazumdar\thanksref{cor}}
\thanks[cor]{email: sumit@physics.arizona.edu}
\author[2]{R.T. Clay}
\author[1]{H. Li}

\address[1]{ Department of Physics, University of Arizona
Tucson, AZ 85721}
\address[2]{Department of Physics and Astronomy and HPC$^2$ Center for
Computational Sciences, Mississippi State University, Mississippi State MS 39762}

\date{\today}
\begin{abstract}
We show that superconductivity is absent within the
$\frac{1}{2}$-filled band triangular lattice repulsive Hubbard model
that has been proposed for organic charge-transfer solids.  We posit
that organic superconductivity is rather reached from a Bond-Charge
Density Wave that either constitutes the insulating state proximate to
superconductivity, or is extremely close in energy to the
antiferromagnetic state, and replaces the latter under pressure. The
Bond-Charge Density Wave can be described within an effective
attractive $U$ extended Hubbard Hamiltonian with repulsive nearest
neighbor interaction $V$.  A first-order transition from the insulating
to the superconducting state occurs within the model with increasing
frustration.

\end{abstract}
\end{frontmatter}

\section{Introduction}

Superconductivity (SC) in the organic charge-transfer solids (CTS) is
proximate to a number of spatial broken symmetry states, including
antiferromagnetism (AFM), the spin-Peierls (SP) state and
charge-ordering (CO).  There are two goals of our presentation. First,
we show from exact diagonalization that SC is absent within the
$\frac{1}{2}$-filled band triangular lattice repulsive Hubbard model
for any parameters. This model Hamiltonian has been proposed to
explain the behavior of $\kappa$-ET and related systems in which SC is
proximate to AFM.  Mean-field calculations find narrow regions of SC
between AFM and PM phases \cite{Powell05a,Kyung06a}. We show from
exact calculations of pair-pair correlations that SC is absent within
the model for any anisotropy or electron-electron (e-e) interaction.
Second, we present evidence that the insulator-superconductor
transition in the CTS is a transition from a Bond-Charge Density Wave
(BCDW).  We point out that the BCDW is identical to the Valence Bond
Solid (VBS) found recently in the EtMe$_3$[Pd(dmit)$_2$]$_2$
compound\cite{Kato06a}.  We argue that the $\frac{1}{4}$-filled band
BCDW/VBS can be modeled by an effective $\frac{1}{2}$-filled band
negative-$U$ Hubbard model with nearest neighbor repulsion $V$. With
increasing frustration a transition from the VBS to SC occurs within
the model. We discuss implications of this work for the CTS.

\section{The repulsive-$U$ Hubbard model}

We consider the Hamiltonian
\begin{eqnarray}
H &=& - t\sum_{\langle ij \rangle,\sigma}(c_{i,\sigma}^\dagger c_{j,\sigma}+ H.c.) 
-t^{\prime}\sum_{[kl],\sigma}(c_{k,\sigma}^\dagger c_{l,\sigma} \nonumber \\
&+& H.c.) +U \sum_{i} n_{i,\uparrow} n_{i,\downarrow}. 
\label{ham}
\end{eqnarray}
In Eq.~\ref{ham} $c^{\dagger}_{i,\sigma}$ creates an electron with
spin $\sigma$ ($\uparrow$, $\downarrow$) on site $i$,
$n_{i,\sigma}=c^{\dagger}_{i,\sigma}c_{i,\sigma}$.  $U$ is the on-site
e-e interaction. The lattice structure is the
conventional square lattice (hopping integrals $t$) with additional
hopping integrals $t^\prime$ across the $x+y$ diagonals for a total of
three bonds per site. The ratio $t^\prime/t$ interpolates from the
square lattice ($t^\prime/t=0$) to the isotropic triangular lattice
($t^\prime/t=1$). In what follows, we express all quantities in units
of $t$.
\begin{figure}[tb]
\centerline{\resizebox{2.75in}{!}{\includegraphics{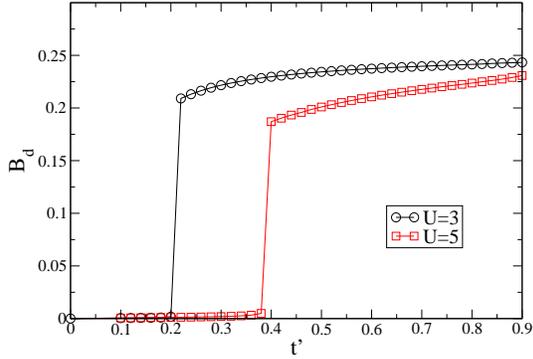}}}
\caption{Diagonal bond order $B_d$ for $U=3$ and $U=5$. The jump
in $B_d$ indicates a transition to a metallic state.}
\label{diagonalbo}
\end{figure}

We report results of exact diagonalizations for $\frac{1}{2}$-filling
within a periodic 4$\times$4 lattice.  For $t^\prime=0$, the ground
state of Eq.~\ref{ham} has long range AFM for any nonzero $U$. With
nonzero $t^\prime$, the ground state is a PM until $U>U_c(t^\prime)$,
when the ground state is an AFM semiconductor. Details of the
insulating and magnetic phases can be found elsewhere
\cite{Mizusaki06a,Clay08b}. Here we focus on the insulator-metal
transition and possible superconducting behavior near the boundary.

The insulator-metal transition can be understood from the behavior of
the bond-order $B(i,j)$ between sites $i$ and $j$, which measures the
probability of charge-transfer and is defined as,
\begin{equation}
  B(i,j)= \sum_{\sigma} \langle c^{\dagger}_{i,\sigma} c_{j,\sigma} +
  H.c.  \rangle \label{Bond-order}
\end{equation}
$B(i,j)$ is expected to be large (small) in the metallic (insulating)
phase. We define the standard pair-creation operator
$\Delta^\dagger_i$,
\begin{equation}
\Delta^\dagger_i= \frac{1}{\sqrt{2}}\sum_{\nu} g(\nu) 
(c^\dagger_{i,\uparrow}c^\dagger_{i+{\bf \nu},\downarrow}
- c^\dagger_{i,\downarrow}c^\dagger_{i+{\bf \nu},\uparrow})
\label{pair}
\end{equation}
In Eq.~\ref{pair} the phase factor $g(\nu)$ is (a) +1 for the four
nearest-neighbor sites $\nu=i+\hat{x}$, $i+\hat{y}$, $i-\hat{x}$ and
$i-\hat{y}$ for $s$-wave pairing; (b) +1,-1,+1,-1 for the same four
sites for $d_{x^2-y^2}$ pairing; and (c) alternates sign over the four
sites $i+\hat{x}+\hat{y}$, $i+\hat{x}-\hat{y}$, $i-\hat{x}-\hat{y}$,
$i-\hat{x}+\hat{y}$ for $d_{xy}$ pairing. We have calculated pair-pair
correlation functions $P(r)=\langle \Delta^\dagger_i\Delta_{i+\vec{r}}
\rangle$ as a function of distance $r$ for all these symmetries as
well as their superpositions. Here we report our results for only the
$d_{x^2-y^2}$ pairing, which is supposed to dominate for $U>0$ and
$t^\prime<1$ \cite{Powell05a,Kyung06a}.

In Fig.~\ref{diagonalbo} we have plotted $B(i,j)=B_d$ for
$j=i+\hat{x}+\hat{y}$ against $t^\prime$ for two different $U$. $B_d$
is a measure of ``metallicity''. Our results correctly demonstrate the
validity of the finite size calculations, with $B_d \sim 0$ for small
$t^\prime$ for both $U=3$ and 5, as expected for the insulating
regime, and $B_d$ showing a sudden jump to a large value at a
$U$-dependent $t^\prime$, where a transition has occurred to either to
the PM phase or to a SC state. We have found similar behavior for
other $U$, $1 \leq U \leq 10$.
\begin{figure}[tb]
\centerline{\resizebox{2.75in}{!}{\includegraphics{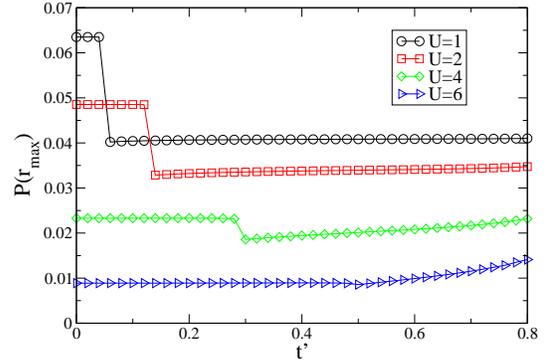}}}
\caption{$P(r_{max}=2\sqrt{2})$ as a function of ${t^\prime}$ for
  $U$=1, 2, 4, 6. Long-range pairing is suppressed by $U$ at all
$t^\prime$.}
\label{correlations}
\end{figure}

A necessary condition for SC within the $U>0$ model is that there
exists at least a range of $U$ within which $P(r)$ at fixed $r$ is
enhanced by $U$. In Fig.~\ref{correlations} we have plotted
$P_{d_{x^2-y^2}}(r=2\sqrt{2})$ against $t^\prime$ for several values
of $U$. As seen in this figure, there occurs a monotonic decrease in
the pair-pair correlation function with $U$ for any $t^\prime$. This
is a clear and convincing proof that any tendency to pairing is
destroyed by $U$.  We have obtained identical results for $s$ and
$d_{xy}$ pairing.

\section{VBS and the effective attractive $U$ model}

An alternate theoretical description of superconducting CTS and
related systems begins from the number of carriers per molecular site
$\rho= \frac{1}{2}$, and emphasizes their $\frac{1}{4}$-filled band
nature \cite{Clay07a,Clay05a,Mazumdar99a,Clay02a,Mazumdar08a}.  The
appropriate theoretical model in this case is the $\frac{1}{4}$-filled
band {\it extended} Hubbard model, with nonzero nearest neighbor
Coulomb repulsion $V$, as well as the intersite and intrasite
electron-phonon (e-p) interactions. We have solved this Hamiltonian
for the quasi-1D (TMTTF)$_2$X \cite{Clay07a}, the zigzag ladder system
DTTTF-Cu(mnt)$_2$ \cite{Clay05a}, the quasi-2D (TMTSF)$_2$X
\cite{Mazumdar99a}and $\theta$-ET$_2$X \cite{Clay02a}.  Two different
CO patterns are possible in these compounds: the Wigner crystal, with
charge-rich sites having only charge-poor sites as neighbors; and the
BCDW/VBS, in which the density wave consists of periodic arrangements
of {\it pairs} of charge-rich sites that form local singlets or
bipolarons. The BCDW results from a co-operative interaction between
the e-p interactions and the AFM coupling, and dominates for realistic
$U \sim 6-8$ and $V \sim 1-3$ \cite{Clay07a,Clay05a,Mazumdar99a,Clay02a,Mazumdar08a}.
\begin{figure}[tb]
\centerline{\resizebox{2.75in}{!}{\includegraphics{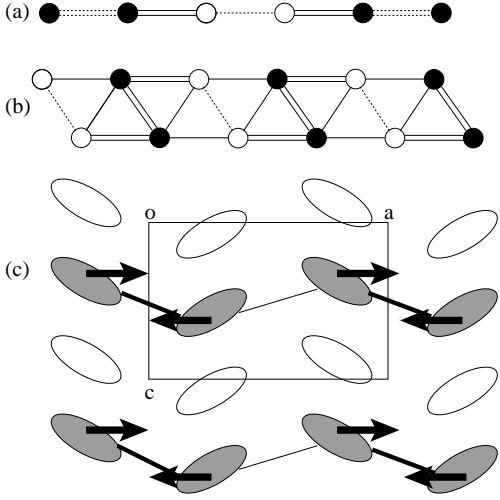}}}
\caption{(a) The ...1100... 1D BCDW state; (b) The 
zigzag ladder BCDW state; (c) horizontal stripe
CO in $\theta-(ET)_2X$  with SP distortion at low temperature 
(see reference \cite{Watanabe07a}). Filled (open) symbols represent
charge-rich (charge-poor) sites in all cases.}
\label{cartoons}
\end{figure}

In Fig.~\ref{cartoons} we have shown the BCDW for the 1D chain, the
zigzag ladder, and the $\theta$-ET triangular lattice with the
dominant horizontal stripe CO. In all these cases the $\rho=
\frac{1}{2}$ BCDW can be modeled as an {\it effective} $\rho=1$
lattice, with each site of the effective lattice composed of {\it
  pairs} of occupied or unoccupied molecular sites. This is most
obvious in the 1D case in Fig.~\ref{cartoons}(a), where the $\rho=\frac{1}{2}$
$\cdots 1100 \cdots$ BCDW (here `1' and `0' denote the charge-rich
and charge-poor sites with actual site charges 0.5 + $\epsilon$ and
0.5 -- $\epsilon$, respectively.)  can clearly be thought of as the
$\rho=1$ density wave $\cdots 2020 \cdots$ (with `2' and `0' having
charges 1.0 + $\epsilon$ and 1.0 -- $\epsilon$, respectively.)

\section{Negative-$U$ extended Hubbard model}

We model such a system by an {\it effective} negative-$U$ extended
Hubbard Hamiltonian \cite{Mazumdar08a},
\begin{eqnarray}
&H& = - t\sum_{\langle ij \rangle,\sigma}(c_{i,\sigma}^\dagger c_{j,\sigma}+ H.c.) 
-t^{\prime}\sum_{[kl],\sigma}(c_{k,\sigma}^\dagger c_{l,\sigma}+\nonumber\\
 &H.c.&) -|U| \sum_{i} n_{i,\uparrow} n_{i,\downarrow} 
+ V\sum_{\langle ij \rangle} n_in_j + V^{\prime}\sum_{[kl]} n_kn_l \nonumber
\label{Hamiltonian}
\end{eqnarray}
All terms have their usual meanings, with
$n_i=\sum_{\sigma}n_{i,\sigma}$; $\langle ... \rangle$ implies
n.n. along the $\hat{x}$- and $\hat{y}$-axes, with hopping integral
and Coulomb repulsion $t$ and $V$, respectively. Similarly, [$\cdots$]
implies neighbors along the ($\hat{x}$+$\hat{y}$)-diagonal, with
$t^{\prime}$ and $V^{\prime}$ as the hopping and Coulomb integrals. We
emphasize that the effective attractive $U$ here originates from the
co-operative interaction between the AFM coupling and e-p interactions
in the original repulsive-$U$ $\frac{1}{4}$-filled band Hamiltonian.
\begin{figure}
\centerline{\resizebox{2.75in}{!}{\includegraphics{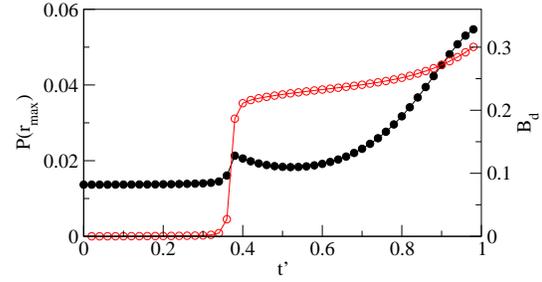}}}
\caption{$B_d$ (open symbols) and $P(r_{max})$ (filled symbols) as a
  function of $t^\prime$ in the negative-$U$ model for $U=-1$,
  $V=V^\prime=1$. Transition to SC occurs at $t^\prime=0.38$.}
\label{pr}
\end{figure}

For small $t^{\prime}$ and $V^{\prime}$, the ground state of
Eq.~\ref{Hamiltonian} is the checkerboard CO, with double occupancies
as neighbors of vacancies and vice versa.  We have done exact
diagonalizations, again on the periodic 4$\times$4 lattice, that finds
a sharp CO-to-SC transition within the model with increasing
frustration. The pair-correlation function here is defined as,
\begin{equation}
P(r)=\frac{1}{N} \sum_j \langle c^{\dagger}_{j,\uparrow} c^{\dagger}_{j,\downarrow}
c_{j+\vec{r},\downarrow} c_{j+\vec{r},\uparrow}\rangle \label{pair-pair} 
\end{equation}
For each combination of $U$, $V$ and $V^{\prime}$ we have found that
$P(r)$ exhibits LRO (to the extent that this can be measured on a
finite lattice) for $t^{\prime}$ greater than a critical value. This
is shown in Fig.~\ref{pr} for the specific case of $U=-1$,
$V=V^{\prime}=1$, where we have plotted both $B_d$ and $P(r_{max}=2\sqrt{2})$
versus $t^{\prime}$.  The behavior of $B_d$ is
similar to that in Fig.~\ref{diagonalbo}, with a semiconducting state
for $t^{\prime}<0.38$. From calculations of charge-charge correlations
and the static structure factor (not shown) we have confirmed that the
semiconducting state is the checkerboard CO. The jump in $B_d$ at
$t^{\prime}=0.38$ is due to the transition to a superconducting state,
as evidenced by the simultaneous jump in $P(r_{max})$. Calculations over
broad ranges of $U$ and $V^{\prime}$ indicate that the transition is
first-order, when the initial CO is strong, and second-order
otherwise.

\section{Significance for the CTS}

We believe that the negative-$U$ Hubbard Hamiltonian with repulsive
$V$ gives the correct insight to understanding the unconventional SC
in the organics, in spite of its limitations.  To begin with, we
believe that the same fundamental mechanism applies to (TMTSF)$_2$X,
ET$_2$X and the anionic [Pd(dimit)$_2$]$_2$ salts. In all these cases
the lattice structure is anisotropic triangular. We believe that the
role of pressure is to increase the interchain coupling and take the
systems towards the more isotropic limit.  We present a brief
discussion of implications of our work below.

1. Although the insulator-to-SC transition in the organics have been
often thought to be an AFM-to-SC transition, as shown in section II,
there is no SC within the triangular lattice $\frac{1}{2}$-filled band
Hubbard model. It is also significant that AFM is missing in the
insulating states of the more isotropic $\kappa$-ET$_2$Cu$_2$(CN)$_3$
\cite{Shimizu03a} and EtMe$_3$[Pd(dmit)$_2$]$_2$ \cite{Kato06a}.  As
we have already pointed out, the experimentally determined VBS in
EtMe$_3$[Pd(dmit)$_2$]$_2$ is charge disproportionated and has the
same charge-densities and intermolecular distances as in our BCDW
\cite{Kato06a}.  We conclude that even when the insulating state is
AFM, pressure-induced frustration leads to a transition to the BCDW,
which in turn becomes superconducting.

2. Partial justification of the above proposed scenario comes from our
recent work on the temperature-dependence of the spatial broken
symmetries in (TMTTF)$_2$X \cite{Clay07a}.  As shown in this work,
here the high temperature CO state is a Wigner crystal, but as the
temperature is lowered and the system enters the SP state, the BCDW
dominates. Similar pressure-induced AFM-to-SP transition here is also
accompanied by a transition from the Wigner crystal CO pattern to the
BCDW \cite{Yu04a}. The implication of these experiments is that the
BCDW is very close energetically to the other broken symmetry states
even when the latter dominate.

3. Very large upper critical field is a characteristic of the
superconducting CTS.  This and the upward curvature of T$_c$
versus magnetic field is a characteristic of superconductors with
local pairing \cite{Micnas90a}.

4. Our theoretical model is a ``superposition'' of the t-J model (the
intersite pairing being driven by AFM correlations) and the so-called
bipolaron model of SC\cite{Micnas90a}.
A critical difference from the standard bipolaron model is the crucial
role of the number of carriers per site $\rho=\frac{1}{2}$, and the resulting
{\it co-operation}, as opposed to competition, between e-e and e-p interactions. 
Both $\rho=\frac{1}{2}$ and frustration are essential ingredients of the
insulator-to-superconductor transition.

5. The bipolaron model has often been criticized because of the
supposedly heavy masses of the bipolarons. This, however, is strictly
true only for the on-site bipolarons and not for intersite
bipolarons. The latter are particularly mobile on triangular lattices
\cite{Hague07a}.

6. The superconducting symmetry within the negative-$U$ extended
Hubbard Hamiltonian can only be s-wave. This, however, need not to be
true within the actual $\frac{1}{4}$-filled band repulsive $U$
extended Hubbard Hamiltonian that provides a more complete description
of the CTS \cite{Clay07a,Clay05a,Mazumdar99a,Clay02a,Mazumdar08a}. The
question of gap symmetry can therefore be resolved only after the true
Hamiltonian is solved. Work is under progress in this direction.

\end{document}